\documentclass[preprint2]{aastex}

\usepackage{rotating}
\usepackage{psfig}

\newcommand{\kms}{{\rm\,km\,s^{-1}}}
\newcommand{\pc}{{\rm\,pc}}
\newcommand{\kpc}{{\rm\,kpc}}

\newcommand{\msun}{{\rm\,M_\odot}}

\newcommand{\Gyr}{{\rm\,Gyr}}

\slugcomment{Submitted to the Astrophysical Journal}
\shortauthors{J. E. Taylor \& Arif Babul}
\shorttitle{Dynamics of Sinking Satellites...}
\begin{document}

\title{The Dynamics of Sinking Satellites Around Disk Galaxies: 
A Poor Man's Alternative to High-Resolution Numerical Simulations}
\author{J. E. Taylor\altaffilmark{1} \& Arif Babul\altaffilmark{1}\\
Department of Physics and Astronomy, Elliott Building, University of Victoria, \\
3800 Finnerty Road, Victoria, BC V8P 1A1, Canada}
\altaffiltext{1}{
Electronic mail: {\tt taylor@uvic.ca, babul@uvic.ca}
}
\begin{abstract}

We have developed a simple yet surprisingly accurate analytic scheme 
for tracking the dynamical evolution of substructure within larger dark halos.
The scheme incorporates the effects of dynamical friction, tidal mass loss and
tidal heating via physically motivated approximations.  Using our scheme, we
can predict the orbital evolution and mass-loss history of individual subhalos 
in detail. We are also able to determine the impact and importance of the 
different physical processes on the dynamical evolution of the subhalos.  
To test and calibrate this model, we compare it with a set of recent 
high-resolution numerical simulations of mergers between galaxies and small 
companions. We find that we can reproduce the orbits and mass-loss rates seen 
in all of these simulations with considerable accuracy, using a single set of 
values for the three free parameters in our model. Computationally, our scheme 
is more than 1000 times faster than the simplest of the high-resolution 
numerical simulations. This means that we can carry out detailed and 
statistically meaningful investigations into the characteristics 
of the subhalo population in different cosmologies, the stripping and 
disruption of the subhalos, and the interactions of the subhalos with 
other dynamical structures such as a thin disk.  This last point is of 
particular interest given the ubiquity of minor mergers in hierarchical models.
In this regard, our method's  simplicity and speed  makes it particularly 
attractive for incorporation into semi-analytic models of galaxy formation.
\hfill\break
\end{abstract}

\keywords{cosmology: dark matter --- galaxies: kinematics and dynamics --- galaxies: interactions --- methods: numerical}

\section{Introduction}

Over the past two decades, observational and theoretical progress have given 
rise to an increasingly detailed picture of how structure forms and evolves
on galactic scales. The currently favored theoretical models are based on the 
concept of hierarchical clustering, in a universe dominated by cold dark 
matter (CDM). Galaxies are embedded within extended halos of dark matter, 
which form through gravitationally induced mergers of 
smaller-scale structure. The evolution of individual dark
matter halos and the formation of galactic structure, if any, 
inside the halos is strongly dependent on the non-linear dynamics of 
gravitational collapse, the stochastic process of merging, and the 
subsequent evolution of the merged substructure. These processes are 
extremely challenging to model theoretically. Consequently the exact role 
of mergers, especially minor mergers, in the 
formation and evolution of galaxies is still poorly understood, even though 
mergers are thought to be an ubiquitous feature of structure formation. 

The dynamics of structure formation have been studied extensively using
numerical simulations. Simulations have been used to investigate the 
properties of populations of cluster, group, and galaxy-sized halos 
in their larger cosmological 
environment (e.g.\ Jenkins et al.\ 1998; Jing \& Suto 1998; 
Governato et al.\ 1998, 1999; Kauffmann et al.\ 1999; Pearce et al.\ 1999; 
Sigad et al.\ 2000), the detailed properties 
of individual halos on scales where their substructure is resolved 
(e.g.\ Ghigna, et al.\ 1998, 1999; Moore et al.\ 1999; Klypin et al.\ 1999a,
1999b; Okamoto \& Habe 1999; Lewis et al.\ 2000; Yoshikawa, Jing \& Suto 2000;
Fukushige \& Makino 2000; Jing \& Suto 2000), 
as well as the outcome of mergers between halos or subcomponents within halos, 
both minor (e.g.\ Quinn \& Goodman 1986; 
Quinn, Hernquist \& Fullagar 1993; Walker, Mihos \& Hernquist 1996; 
Huang \& Carlberg 1997; Vel{\'a}zquez \& White 1999) 
and major (e.g.\ Barnes 1998; Naab, Burkert \& Hernquist 1999).
Existing simulations do not yet have the dynamic range to 
explore these different scales simultaneously, however, leading to a 
fragmented treatment of the subject. 
Moreover, numerical simulations suffer
from several other disadvantages: (1) they are very expensive computationally; 
(2) the detailed statistical properties of substructures within a halo 
may depend sensitively on the scheme used to identify them; (3) the evolution
of these objects may be influenced in complex ways by the number of particles 
used to resolve them, as well as other numerical effects such as finite force 
resolution (e.g.\ Ghigna et al.\ 1999; Knebe et al.\ 2000). 

The above limitations are particularly frustrating, given the recent
suggestions that hierarchical models may be failing
to match observations on galactic scales. One
apparent discrepancy is in the number of satellites
expected to have survived in a Milky Way-sized halo versus the observed number
of satellites around the two large galaxies of the Local Group (Klypin
et al.\ 1999b; Moore et al.\ 1999; Bullock, Kravtsov \& Weinberg\ 2000). 
This excess structure may be
implicated in several other problems, including the small disk sizes
produced in hydrodynamic simulations (e.g.\ Navarro \& Steinmetz 2000), 
and the question of disk
survival against heating in minor mergers (T\'{o}th \& Ostriker 1992;
Kauffmann \& White 1993; Lacey \& Cole 1993; Navarro, Frenk \& White 1994;
Moore et al.\ 1999).
To resolve these issues,
it is necessary to consider separately the effects of background cosmology,
the power spectrum of density fluctuations, and most importantly 
the nature and dynamics of substructure within galactic halos. 
Accomplishing this goal numerically would entail exploring a large parameter 
space, via ultra-high resolution simulations. This proposition is, however, 
prohibitively expensive at present.

An alternate approach to studying structure formation is to use 
semi-analytic (SA) methods, which combine analytic theory and 
numerical results. Semi-analytic models of galaxy formation 
(e.g.\ Kauffmann, White \& Guiderdoni 1993; Cole et al.\ 1994; 
Somerville \& Kolatt 1999) generate random realizations of merging 
(or ``merger histories'' or ``merger trees'') between halos based on 
Press--Schechter statistics (Press \& Schechter 1974). The formation and evolution of galactic 
structure within the halos is then governed by a set of prescriptions 
which attempt to describe the effects of merging, hydrodynamics, shocks, 
dissipation, star formation and feedback.

Semi-analytic models of galaxy formation have thus been used to study 
the galaxy
luminosity function, the Tully--Fisher relation, the morphology--density
relation and other global properties of galaxy populations 
(see Somerville \& Primack 1999 for a recent review, as well as
Bullock et al. 2000 for more recent work on galactic satellites). 
SA methods have the advantage of being extremely fast compared to fully 
numerical simulations, and although the results depend on the 
prescriptions adopted (Benson et al.\ 1999), these are, at least in 
principle, transparent, so that it is easy to test the consequences 
of modifying them. SA methods are therefore useful for exploring the 
relative importance of the various ingredients of the galaxy formation 
model ---
the background cosmology and the shape of power spectrum, the 
density profiles of dark matter halos, the non-linear dynamics of merging, 
the gas and radiative physics, star formation and feedback algorithms --- 
in determining the appearance of present-day galaxies. 

Until recently, however, SA models of galaxy formation have focused on star 
formation and dissipative processes, and have included only a simplified 
description of merging, or even ignored the impact of merging altogether. 
In Diaferio et al.\ (2000), for example, nearly equal-mass mergers result 
in the destruction of the disk, but the impact of uneven-mass mergers on the 
galaxy embedded in the more massive halo is ignored. This makes it difficult 
to ascertain whether, for example, morphology-related results are real or 
merely artifacts of the oversimplified prescription. It also makes SA results 
hard to relate to studies of the dynamical evolution of individual galaxies, 
whether analytic (e.g.\ Dalcanton, Spergel \& Sommers 1997; 
Mo, Mao \& White 1998) or numerical (e.g.\ Vel{\'a}zquez \& White 1999).

Studying various issues arising from hierarchical clustering scenarios, 
including the distribution of satellites around galactic systems, the impact 
of these satellites on a thin disk, and the formation of the stellar halo 
from tidal debris, requires a method for determining the evolution of the 
satellites which synthesizes and generalizes the results of existing numerical 
studies, without resorting to expensive ultra-high resolution simulations. 
The method should take into account the internal structural properties of a 
satellite, its orbital parameters and the details of its interaction with 
the main galaxy.

We have developed a simple analytic scheme that addresses this need
and complements existing semi-analytic and numerical models 
of galaxy formation. 
We consider subhalos within a galaxy halo individually, following 
their orbits and accounting for dynamical friction, mass loss and heating 
using analytic expressions.
This approach will allow us to carry out detailed studies of the 
properties of subhalos for a variety of cosmologies and power spectra.
Since the physical processes underlying our scheme are modeled explicitly,
we can determine their relative effects on subhalo evolution directly.
Finally, as our model allows us to generate large numbers of realizations 
at little cost, we can do all of the above in a statistically meaningful 
manner. We also note that although our method is described in the context of 
galaxy-sized halos, it is completely general and can also be used to study
the evolution of substructure in clusters, for instance.

In this paper, we outline our dynamical model for the evolution 
of substructure. In section 2, we give a brief synopsis of the
previous investigations of the dynamics of satellites merging with a 
larger galactic system, and then describe the theory underlying
our scheme for following orbital decay, mass loss, and the tidal disruption of 
subhalos. In section 3, we test and calibrate our model by comparison with
recent high-resolution numerical simulations of sinking 
satellites by Vel{\'a}zquez \& White (1999) (VW hereafter). 
In section 4, we explore how tidal heating, the form of the galactic potential
and the subhalo mass profile affect mass loss and orbital decay. 
We summarize our results in section 5. In subsequent papers we will 
combine our model of dynamical evolution with semi-analytic merger trees 
to study the halo substructure and disk heating produced in a galaxy halo
by multiple mergers with cosmologically realistic satellites on
representative orbits. 

\section{Dynamics of Merging Substructure}

\subsection{Background}

The first detailed study of the dynamics of satellites evolving within a
halo containing a disk galaxy 
was carried out by Quinn \& Goodman (1986). The study was subsequently 
improved upon by Quinn et al.\ (1993).
These authors investigated the problem using numerical simulations and found 
that while dynamical friction could not account for the anisotropy in 
the orbits of satellites around spiral galaxies observed by Holmberg (1969), 
the
decay times were indeed short for large satellites, as expected from 
analytic arguments, although there was some variation depending on
the details of the orbits. Prograde orbits close to the plane of the disk tended
to decay the fastest. The studies considered only satellites on initially circular
orbits and found that the sinking of the satellites took place in two steps:
(1) A relatively slow decay of the orbital radius largely dominated by loss of altitude
with respect to the disk; (2) a rapid decay in the radius once the satellite is 
in the disk plane. The results also suggested that the massive satellites would 
heat the disk appreciably, although the effect was complex and depended on the 
orbit and the internal structure of the satellite. Finally, the authors noted that 
noise in the simulations is a significant factor and that even in simulations 
with $\sim 500,000$ particles, the noise would make it difficult to study the 
dynamics of the system over the Hubble time.

In the period between Quinn \& Goodman (1986) and Quinn et al.\ 
(1993), T\'{o}th \& Ostriker (1992) (TO hereafter) 
studied the effects of minor mergers on the 
structure of the Milky Way's 
disk using a semi-analytic model for the evolution of the satellites, and
the disk heating they produced. The orbital energy of a satellite lost
through dynamical friction was assumed to go into heating the disk locally.
Comparing their predictions to the observed scale height of the disk and to 
the local value of Toomre's stability parameter $Q$, they concluded that the 
Milky Way could not have accreted more than 4\%\ of its mass interior to the 
solar circle in the past 5\Gyr. 

Several more recent studies have sought to evaluate the results of 
TO. 
Walker, Mihos \& Hernquist (1996) (WMH hereafter) and Huang \& Carlberg (1997) 
(HC hereafter) used numerical
simulations, improving upon previous work by including a responsive halo.
WMH opted to use a very large number of particles 
to reduce
numerical noise and in turn, had to start their satellites fairly close in,
at a distance of $21\kpc$ from the center of the galaxy. They found that 
the evolution of the satellite was similar to that described by Quinn \& Goodman 
(1986) and Quinn et al.\ (1993). As for the disk, they found that 
accretion of a satellite with 10\%\ of the disk mass would result in a significant
thickening of the disk at the solar radius. HC, on the
other hand, considered satellites, starting out at larger distances, with masses 
of 10--30\%\ of the disk mass. The internal structure of the satellites
consisted of a small tightly bound core embedded in an extended low-density 
envelope 
containing most of the mass. As a result, the satellites tended to be 
disrupted by tidal forces before they could heat the disk significantly. 
For completeness, we note that Weinberg (1995, 1998) and Sellwood, 
Nelson \& Tremaine (1999) have also studied the effects of satellite-disk
interactions, the focus of these studies being on the response of the 
disk.

The most recent numerical study of satellite-disk interactions is that
of VW, who studied mergers involving several
different satellites, intermediate in density between those of 
TO and those of 
HC, on various different orbits. They confirmed that 
Chandrasekhar's formula (Chandrasekhar 1943) gives a useful approximation 
to the drag force exerted by the halo on the satellite 
provided the Coulomb logarithm is adjusted separately for each orbit.
VW also found that the response of the disk depends partly on the orientation
of the satellite orbit, 
prograde encounters tending to heat the disk preferentially, while retrograde 
encounters tend to tilt it. They concluded that TO overestimated 
the magnitude of disk heating by a factor of 2--3 overall.

Although much progress has been made in studying the effect
of minor mergers on galactic structure, it is hard to determine the 
cosmological implications of merger simulations given the lack of clear, 
consistent, and robust results across the different studies. 
This is, first and foremost, due to the fact that each 
study has considered satellites with different internal properties. 
Structural characteristics such as the satellite's density profile 
affect the rate of mass loss and therefore the evolution of the satellite's 
orbit, not to mention the response of the disk to the satellite. In addition, the
satellites in the different studies were started at different radii, and 
early studies did not include the dynamical friction
produced by the halo. On a more
practical level, the simulations were subject to
numerical effects such as finite force resolution, shot noise, relaxation, 
and artificial heating from interactions between particles of different masses,
which differed from study to study. Finally, all studies prior to 
that of VW focused on satellites on circular or 
nearly circular orbits, making it difficult to generalize their results 
to satellites accreted by galaxies in self-consistent cosmological settings. 
To analyze this prior body of numerical work, and overcome the limitations 
mentioned previously, requires an alternative method which follows the 
relevant physical processes explicitly, and can generate many realizations
of merging at little computational expense.

\subsection{The Semi-Analytic Model}

To model the evolution of a subhalo as it merges with a larger halo, 
we treat it as a spherically symmetric satellite, with structural properties 
that change over time. The structure of the satellite is fully specified at 
any time by its mass, the form of its density profile, its initial 
core or scale radius, a limiting outer radius, and the amount of heating 
it has experienced. The density profile is initially set to a standard form
such as a King model, or any one of several common analytic density profiles. 
If the satellite experiences tidal heating, however, its 
density profile may change. We do not track these changes 
explicitly (although they may be calculated from our other results), 
as we are only concerned with changes in the mean density of the satellite
within its limiting radius in our description of heating and mass loss.

To determine the satellite's orbit,
we ignore its spatial extent to first order, and calculate the trajectory of 
a point particle with the same total mass, moving in the gravitational 
potential of the halo-galaxy system. This approximation is entirely sufficient
as long as the scale of the satellite's orbit is larger than the satellite
itself; if the satellite falls into the central region of the potential, we 
consider it to be disrupted in any case. The background potential is taken to 
be static in the present study, although in general it can be allowed to vary 
in a self-similar way in our code.

To account for dynamical friction, the response of the halo and disk to the 
satellite, we impose a drag force on the satellite which we calculate using 
Chandrasekhar's formula. As mentioned above, we also adjust the satellite's 
total mass and modify its internal structure in response to tidal stripping 
and tidal heating, respectively, during the course of its orbit. The satellite 
is considered disrupted when it has been stripped down to its core radius, or 
when it falls into the central region of the potential. Each of these 
processes is described separately below.

\subsubsection{Dynamical Friction}

Chandrasekhar (1943) showed that a massive particle moving through 
a distribution of background particles will generate a wake. 
The collective gravitational force from the wake will act
back on the massive particle, causing a drag force know as dynamical friction. 

Dividing the background potential into its two kinematically distinct
components, we use Chandrasekhar's formula to 
estimate the dynamical friction exerted by the halo/bulge system 
and by the disk on an orbiting satellite:

\begin{eqnarray}\label{dragforce}
\hskip-0.42truecm{{\mathbf F}_{\rm df}}&=&{\mathbf F}_{\rm df,halo}\ +\ {\mathbf F}_{\rm df,disk}\nonumber\\
&=&-\ {4\pi {G^2} M_{\rm sat}^2}\nonumber\\
&&\hspace{0.2truecm}\ \ \ \times \hspace{-0.1truecm}\sum_{i = {\rm h,d}} \rho_i(<V_{{\rm rel},i}) \ln\Lambda_i {{{\mathbf V}_{{\rm rel},i}} 
\over {{\left|{\mathbf V}_{{\rm rel},i}\right|^3}}}
\end{eqnarray}
where\ \ ${\mathbf V}_{{\rm rel,h}} = {\mathbf V}_{\rm sat},\ {\mathbf V}_{{\rm rel,d}} = {\mathbf V}_{\rm sat} - {\mathbf V}_{\rm rot},$\hfill\break

{$\rho_i(<V_{{\rm rel},i}) = \rho_i({\mathbf r})\left[{\rm erf} (X_i) - X_i\ {\rm erf}^\prime (X_i)\right],$}\hfill\break

\noindent{and $\ \ X_i \equiv {\left|{\mathbf V}_{{\rm rel},i}\right|} /({\sqrt 2}\sigma_i).$}

Here $M_{\rm sat}$ is the mass of the satellite,
${\mathbf r}$ is its position,
${{\mathbf V}_{\rm sat}}$ is its velocity, 
${{\mathbf V}_{\rm rot}}$ is the local circular velocity of the disk, 
$\rho_{\rm h}$ is the local density of the spherical (halo/bulge) component,
$\rho_{\rm d}$ is the local density of the disk, 
$\ln\Lambda_{\rm h}$ and $\ln\Lambda_{\rm d}$ are the Coulomb logarithms for the halo/bulge and the disk,
and $\sigma_{\rm h}$ and $\sigma_{\rm d}$ are the one-dimensional velocity dispersions
of the halo/bulge particles and the disk particles 
respectively (Binney \& Tremaine 1987). 

The derivation of this formula assumes a massive point particle,
moving through an infinite, homogeneous background 
of much lighter particles with an isotropic Maxwellian velocity distribution 
of zero mean. Numerous detailed studies of satellite dynamics 
(Weinberg 1986; Cora, Muzzio \& Vergne 1997; Bontekoe \& van Albada 1987;  
van den Bosch et al.\ 1998; VW; Colpi, Mayer \& Governato 1999),
have shown the formula to be more widely applicable, however, in the 
sense that it gives a useful approximation to the drag force 
on an extended satellite in a finite halo-galaxy system provided 
that the Coulomb logarithms are adjusted appropriately. In general,
Binney \& Tremaine (1987) suggests that Chandrasekhar's formula will 
be fairly accurate provided that the mass of the satellite does not exceed
20\%\ of the mass of the larger system, and that the orbit of the satellite
lies neither outside the larger system nor completely within its core.

The argument of the Coulomb logarithm can be expressed as
$\Lambda=b_{\rm max}/b_{\rm min}$,
where $b_{\rm max}$ and $b_{\rm min}$ are measures of the maximum and the
minimum impact parameters of the background particles contributing to the wake. 
For a finite background system, $b_{\rm max}$ is conventionally taken to be
the characteristic scale of the system.
Possible choices for a spherically symmetric system 
include the half-mass radius of the system (e.g. Quinn
\& Goodman 1986), the distance over which the background density 
changes by a factor of two (Binney \& Tremaine 1987), and the tidal radius of 
the halo or the distance between the satellite's position and the center
of background system (Colpi et al.\ 1999). 

The value of $b_{\rm min}$ is equally ambiguous. 
For a point mass satellite, $b_{\rm min} \equiv G(M_{\rm sat}+m)/V^2$, 
where $m$ is the mass of the background particles and $V$ is 
a velocity ``typical'' of the encounter, such as the r.m.s. 
velocity of the background particles (Chandrasekhar 1943), or their mean 
velocity relative to the satellite (Binney \& Tremaine (1987).
For extended satellites, White (1976) derived an expression for $b_{\rm min}$ 
which is approximately equal to $0.2\,r_{\rm t}$ (or very roughly the half-mass
radius) for a wide range of King profiles,
while Quinn \& Goodman (1986) take $b_{\rm min}$ to be the larger of 
the half-mass radius of the satellite and the point-mass value
$G(M_{\rm sat}+m)/V^2$, with $V$ taken as the mean velocity of the
satellite with respect to the background particles.

The choice of an appropriate Coulomb logarithm to describe friction
from the disk, and more generally the applicability of Chandrasekhar's
formula to an inhomogeneous distribution of background particles, is
even less clear. Maoz (1993) 
and Dom\'{\i}nguez-Tenreiro \& G\'{o}mez-Flechoso
(1998) derived formulae for the magnitude
of the energy loss produced by an arbitrary distribution of uniform
velocity dispersion, but could not specify the direction of the corresponding
frictional force. It is also possible to calculate dynamical friction
for a uniform background of particles with an ellipsoidal velocity distribution
(Binney \& Tremaine 1987), in which case the frictional force is strongest
in the direction of the smallest principal axis of the distribution. 
Since disk friction is of secondary importance compared with halo friction, 
however, we shall limit ourselves to using Chandrasekhar's formula 
to calculate its approximate direction and magnitude.

As noted above, in calculating dynamical friction we have 
and taken the density of the satellite's wake to be constant,
and equal to the background density at the centre of the satellite.
Since the wake has a finite extent, this approximation may
result in errors in the drag force if the background density changes 
over small scales. This is likely to occur,
for example, when the satellite is in the plane disk, because of the
latter's small vertical scale height.
To correct for this, the disk density $\rho_{\rm d}$ used in 
equation (\ref{dragforce}) should be smoothed in the vertical 
direction. In principle, the smoothing length ought to be 
related to the characteristic scale of the wake or the satellite; 
however, this would mean using a different smoothing scale for each 
individual satellite.
At present this level of complication does not seem warranted,
nor would it fully account for finite-size effects 
(see Dom\'{\i}nguez-Tenreiro 
\& G\'{o}mez-Flechoso 1998 for a more detailed discussion of the drag force
on extended objects). We have therefore chosen to smooth the disk density in 
the vertical direction by a fixed length corresponding to two times the disk 
scale height, noting that this smoothing 
length is on the order of the half-mass radius of satellites with 
masses in the range where dynamical friction has a substantial effect.

Given the uncertainties associated with calculating the value of the 
Coulomb logarithm, on the one hand, and the fact that, on the other hand, 
Chandrasekhar's formula with an appropriately adjusted Coulomb logarithm 
gives an excellent approximation to the drag force seen in numerical studies, 
we will treat $\ln\Lambda_{\rm h}$ and $\ln\Lambda_{\rm d}$ as free parameters.
We adjust their values in order to locate a point in the 
$\ln\Lambda_{\rm h}$--\,$\ln\Lambda_{\rm d}$ plane for which our semi-analytic 
results provide the best overall match to a series of fifteen satellite 
simulations carried out by VW, the most detailed of such simulations carried 
out to date. One could in principle tune the logarithms to fit each simulation 
separately. Instead, we prefer to identify a single set of values that works 
well for all the orbits, since we would like to determine values of 
$\ln\Lambda_{\rm h}$ and $\ln\Lambda_{\rm d}$ that can be used in more 
general studies of satellite-disk interactions.

\subsubsection{Mass Loss}

The magnitude of the drag force on a satellite due to dynamical 
friction depends on its mass. A satellite with a finite extent will lose mass 
as it orbits within the halo-galaxy system, and as it 
loses mass, the drag force it experiences will decrease. 
As a result, mass loss 
can significantly alter the dynamics of the satellite. To account for this, 
we need to estimate the amount of the mass that remains bound to the satellite 
throughout the course of its orbit. 

Material becomes unbound from the satellite through the action of tidal
forces. Slowly varying and rapidly varying tidal forces will affect the 
satellite differently. In a slowly varying system, material outside some 
limiting, ``tidal'', radius will be stripped from the satellite, 
while in a rapidly
varying system, material throughout the satellite will be tidally heated.
These two regimes have been studied previously by making the approximation 
that the system is static in the first case
(that is a satellite on a circular orbit), 
or that the satellite undergoes a very short perturbation but is
otherwise isolated in the second case (the impulse approximation). 
In this section, we will consider tidal 
stripping on general orbits. Tidal heating will be treated separately 
in the section that follows.

For a satellite on a circular orbit of radius $r$ within a spherically 
symmetric mass distribution, the combined potential of the entire system is
static in the rotating frame. In this case, we can identify the tidal radius 
with the distance to the saddle point in the potential interior to the 
satellite's orbit, since this is the point where the radial forces on a test 
particle cancel out (von Hoerner 1957; King 1962; Binney \& Tremaine 1987). 
The distance $R_{\rm t}$ from the satellite center to this point is:
\begin{eqnarray}\label{Jacobi} 
R_{\rm t} \simeq {\left({GM_{\rm sat} \over {\omega^2 - {\rm d}^2\Phi /{\rm d}r^2}}\right)}^{1/3}
\end{eqnarray}
(King 1962), where $M_{\rm sat}$ is the mass of the satellite, 
$\omega$ is its angular velocity
and $\Phi$ is the potential of the main system. 

This estimate of the tidal radius is only formally valid when 
$M_{\rm sat}$ is much smaller than the mass of the main system, 
$R_{\rm t}$ is much smaller than the orbital radius, and the 
satellite is corotating at its orbital frequency. 
Even under these restricted assumptions, 
the mass inside $R_{\rm t}$ is only approximately equal to the bound mass, 
because there exist orbits that extend beyond $R_{\rm t}$ but remain bound 
to the satellite (Binney \& Tremaine 1987, and references therein). 
Furthermore, even in this simple case, the tidal boundary is not spherical 
and thus the use of expression (\ref{Jacobi}) is approximate.\footnote{ 
Innanen, Harris \& Webbink (1983), for instance, calculate a slightly different value 
for the limiting radius based on the length of the short axis of a tidally 
distorted satellite.} 

General satellite orbits are neither circular, nor is the external potential 
in which they are moving necessarily spherical. We can still use equation 
(\ref{Jacobi}) to define an instantaneous tidal limit for the system,
where $\omega$ is now the instantaneous angular velocity of the satellite.
For non-circular orbits, or orbits out of the plane of the disk, $R_{\rm t}$ 
changes with time, however, and mass outside $R_{\rm t}$ will become unbound 
as a result of successive accelerations over the course of the orbit, rather 
than being 
stripped immediately. While equation (\ref{Jacobi}) represents a steady-state
solution to mass loss, the characteristic timescale for transient changes in
mass on general orbits should be the orbital period. To model this type of mass 
loss, we therefore assume that the satellite mass beyond $R_{\rm t}$ is lost 
over the course of one orbital period, and scale the mass loss in each
timestep accordingly.

In calculating ${\rm d}^2\Phi /{\rm d}r^2$, we will
average over the asphericity of the potential due to the 
disk component. We set:
\begin{equation}
{{{\rm d}^2\Phi} \over {{\rm d}r^2}} = {{{\rm d}^2\Phi_{\rm sph}} \over {{\rm d}r^2}} = {{\rm d}\over{{\rm d}r}}\left({{-GM(<r)}\over{r^2}}\right)\,.
\end{equation}
where $\Phi_{\rm sph}$ is the potential produced by a spherically symmetric
distribution with mass $M(<r)$ interior to $r$. 
This will be very close to the radial gradient of the actual force on the 
satellite when it is far from the disk, or when it is in the plane of the disk.
Only when 
the satellite is close to the disk, but on an inclined orbit, will the true 
gradient differ substantially from this value, and in practice we expect 
tidal shocking to dominate the physics of the mass loss in these cases.  

We can write 
the stripping condition in terms of densities; the tidal limit occurs
at the radius $R_t$ within which the mean density satellite 
$\overline{\rho}_{\rm sat}$ 
exceeds the 
density of the galaxy interior to its position, 
$\overline{\rho}_{\rm gal},$
by a factor $\eta$:
\begin{equation}\label{overd}
\overline{\rho}_{\rm sat} (< R_{\rm t}) = \eta \overline{\rho}_{\rm gal}(<r)\,,
\end{equation}
with
\begin{eqnarray}
\eta &\equiv& {{\overline{\rho}_{\rm sat} (< R_{\rm t})}\over{\overline{\rho}_{\rm gal}(< r)}}\ \ =\ \ {{r^3} \over {R^3_{\rm t}}}\ {{M_{\rm sat}} \over {M(<r)}}\nonumber\\
&=& \left({{r^3} \over {G M(< r)}}\right) \left(\omega^2 - {{{\rm d}^2\Phi} \over {{\rm d}r^2}}\right)\nonumber\\
&=& \left({{\omega^2} \over {\omega_c^2}} - {1 \over {\omega_c^2}}{{{\rm d}^2\Phi }\over {{\rm d}r^2}}\right),
\end{eqnarray}
where $\omega$ is the instantaneous angular velocity of the satellite, 
and $\omega_c$ is the angular velocity of a circular orbit of radius $r$. 

This leads to a particularly simple algorithm for
stripping satellites. First, we divide the satellite's 
orbital path into discrete sections corresponding to fixed timesteps.
In each timestep, we determine the tidal radius of the satellite using 
equation (\ref{overd}). Of the material outside this radius, we remove 
a fraction $\Delta t / t_{orb}$, where $\Delta t$ is the length of the timestep 
and $t_{orb} = 2 \pi / \omega$ is the orbital period, which we assume to be
the characteristic timescale for mass loss. Finally,
we treat the satellite as disrupted and set its bound mass to zero when 
the tidal radius becomes smaller than the core radius of the profile,
although by this point it has normally lost so much mass that the exact 
disruption
criterion is unimportant in practice. As mentioned above, we also 
treat satellites which have fallen into the core of the bulge as disrupted, 
to avoid instabilities in the orbital calculation.

\subsubsection{Tidal Heating}

Whereas a steady or slowly varying tidal field will result in the stripping of
loosely bound mass, a rapidly changing gravitational field, caused, 
for example, by fast encounters with the galactic disk or bulge, will induce 
gravitational shocks that can add energy to the satellite, changing its 
structure and accelerating mass loss
(Ostriker, Spitzer \& Chevalier 1972; Spitzer 1987; 
Kundi\'{c} \& Ostriker 1995; 
Gnedin \& Ostriker 1997, 1999; Gnedin, Hernquist \& Ostriker 1999). 

Tidal heating from shocks changes both the mean and the dispersion
of particle energies within the satellite; to model heating fully requires 
a Fokker--Planck code to track the changing distribution function. In 
keeping with the simple method for estimating tidal mass loss developed above, 
we will derive a first-order correction for tidal heating, and scale it
to match the mass loss rates seen in the simulations. To do this we first
identify rapid shocks by comparing the shock timescale to the satellite's 
internal orbital period. 
Specifically, we heat the satellite only when 
$t_{\rm shock} < t_{\rm orb,sat}$, where $t_{\rm shock} 
\equiv (t_{\rm sh,d}^{-1} + t_{\rm sh,b}^{-1})^{-1}$ is an average
of the disk and bulge shock times $t_{\rm sh,d} = z/V_{z,{\rm sat}}$ and 
$t_{\rm sh,b} = r/V_{\rm sat}$, weighted so that the shorter time dominates,
and $t_{\rm orb,sat} = 2 \pi r_{\rm h} / V_c (r_{\rm h})$ is the orbital 
period of the satellite at its half mass radius. 
This corresponds to the range of shock timescales considered by 
Gnedin \& Ostriker (1999). Over the course of each rapid shock, 
we then calculate the first-order change in energy within the satellite, 
and estimate how the satellite's density profile will change as a result 
of this energy input.

In order to model the effect of tidal heating on the satellite, we use the 
formalism of Gnedin et al.\ (1999). Consider an element of 
unit mass, with coordinates ${\mathbf x}$ with respect to the satellite center. 
In the impulse approximation, tidal acceleration acting over the course of a 
{\it rapid} encounter, of duration $t$, will induce a velocity change:
\begin{eqnarray}
\Delta {\mathbf V} = \int^t_0 {\mathbf A}_{\rm tid}(t')dt'
\end{eqnarray}
relative to the satellite's center of mass, where ${\mathbf A}_{\rm tid}$
is the tidal acceleration. The resulting first-order change in its energy 
is simply equal to the work done by the tidal forces:
\begin{eqnarray}
\Delta E_1 (t)\hspace{-0.2truecm}&=&\hspace{-0.2truecm}W_{\rm tid}(t) = {1 \over 2}\Delta V^2 \nonumber\\
&=&\hspace{-0.2truecm}{1 \over 2}\int^t_0 {\mathbf A}_{\rm tid}(t')dt' \,{\mathbf \cdot} \hspace{-0.1truecm}\int^t_0 {\mathbf A}_{\rm tid}(t'')dt''.
\end{eqnarray}
If we divide the shock into a series of $n$ discrete time steps of length $\Delta t$, then the work done is:
\begin{eqnarray}
W_{\rm tid}(t_n) = {1 \over 2}\,\Delta t^2\left[ \sum^{n-1}_{i = 0} {\mathbf A}_{\rm tid}(t_i)\,{\mathbf \cdot}\hspace{-0.1truecm}\sum^{n-1}_{j = 0} {\mathbf A}_{\rm tid}(t_j) \right].
\end{eqnarray}
In going from $t_n$ to $t_{n+1}$, the energy change in a single timestep is:
\begin{eqnarray}\label{dW1}
\lefteqn{\Delta W_{\rm tid}(t_n \rightarrow t_{n+1})} \nonumber\\
&\hspace{-0.6truecm}\lefteqn{=}&\hspace{-0.3truecm}{1 \over 2}\,\Delta t^2 {\mathbf A}_{\rm tid}(t_n)\,{\mathbf \cdot}\hspace{-0.1truecm}\,\left[ 2\sum^{n-1}_{i = 0} {\mathbf A}_{\rm tid}(t_i) + {\mathbf A}_{\rm tid}(t_n)\right]. 
\hspace{0.7truecm}
\end{eqnarray}

If the satellite is sufficiently small, we can express the tidal force
in terms of the gradient of the gravitational force due to the external 
potential, evaluated at at the center of the satellite:
\begin{eqnarray}
{\mathbf A}_{\rm tid}(t) = {\mathbf x}(t)\,{\mathbf \cdot}\,[{\mathbf \nabla} {\mathbf g}]_{({\mathbf x} = 0)} = g_{a,b} x_b(t)\,{\mathbf e}_a\,,
\end{eqnarray}
where 
${\mathbf g}$ is the external gravitational field,
$g_{a,b} = \partial g_a/\partial x_b$ evaluated at ${\mathbf x} = 0$, 
${\mathbf e}_a$ is the unit vector in the $x_a$-direction
and repeated indices $a$, $b$ indicate summation over the three Cartesian 
coordinates.
Thus taking the dot product in equation (\ref{dW1}) 
and averaging over a sphere of radius 
$r$ gives:
\begin{eqnarray}\label{dW2}
\lefteqn{\Delta W_{\rm tid}(t_n \rightarrow t_{n+1})}\nonumber\\
&=&{1 \over 6}\,r^2\,\Delta t^2\,\Biggr[\,2\,{g}_{a,b}(t_n)\sum^{n-1}_{i = 0} {g}_{a,b}(t_i)\nonumber\\
& &\ \ \ \ \ \ \ \ +\ \ \ g_{a,b}(t_n)g_{a,b}(t_n) \Biggr]
\end{eqnarray}
with eighteen terms from the summation over $a$ and $b$, 
where we have used the fact that 
$$\langle x_i x_j\rangle = {1 \over 3} r^2 \delta_{ij}$$ averaged over a sphere. 
(As explained below, this is only strictly true if the shock is rapid, 
so that $x_i(t_0) \simeq x_i(t_j) \simeq x_i(t_n)$ for all $t_j < t_n$.)

There are two important corrections to equation (\ref{dW2}). 
First, our calculation is based on the impulse approximation, that is the
mass element is assumed to remain stationary over the course of the shock. 
This approximation is expected to break down in the central regions of the 
satellite where the dynamical timescales can be comparable to or 
shorter than the shock duration. In these regions the effects of the shock 
will be greatly reduced. 
To account for this, we adjust the heating
during rapid shocks using the first-order adiabatic correction discussed by
Gnedin \& Ostriker (Gnedin \& Ostriker 1999, and references therein): 
\begin{equation}\label{adc}
\Delta E_1 = A_1(x)\Delta E_{1,{\rm imp}}\ ,
\end{equation}
where
$A_1(x) = (1 + x^2)^{-\gamma}$ and $x$,
the adiabatic parameter, is the ratio of the shock duration $t_{\rm shock}$ 
and the orbital period of the satellite at its half-mass radius $t_{\rm orb,sat}$.
Since most of the heating in our model comes from fairly rapid disk 
shocks, we use a value
of 5/2 for $\gamma$, in keeping with the results of Gnedin \& Ostriker (1999).

Second, heating also leads to a change in the internal velocity dispersion 
of the satellite, as discussed by Kundi\'{c} \& Ostriker (1995). Both the overall 
energy gain and the increase in the dispersion will cause some of the mass to 
become unbound. In keeping with the simplicity of our semi-analytic approach, 
we only compute the first-order gain in energy but account for the higher-order 
effects through the introduction of a heating coefficient, $\epsilon_{\rm h}$,
that we will adjust to yield reasonable overall matches to the VW simulations:
\begin{equation}
\Delta E = \epsilon_{\rm h} \Delta E_1 
\end{equation}
where
$$\Delta E_1 = A_1(x)\Delta E_{1,{\rm imp}}  = A_1(x)\Delta W_{\rm tid}\ .$$
Kundi\'{c} \& Ostriker (1995) estimate that the second-order heating term has
an effect comparable to or greater than that of the first-order term, so we
expect $\epsilon_{\rm h}$ to be greater than 2. From the disruption
timescale arguments in Gnedin \& Ostriker (1997), for instance, we might
expect that $\epsilon_{\rm h} \simeq 7/3$. The value of $\epsilon_{\rm h}$ 
used in practice, however, will also depend on the shocking criterion and 
the adiabatic parameters, as explained below.

To determine how heating affects the satellite, we assume that the change 
in its mass distribution does not involve shell crossings and that the potential energy of a mass element remains proportional to its total energy (as it 
would in virial equilibrium). The total energy $E(r)$ of a mass element at a 
radius $r$ will thus be proportional to $-1/r$, so that an injection of 
energy $\Delta E(r)$ will result in a change in the radius 
$\Delta r \propto \Delta E(r)\ r^2$. In the absence of shell crossings, 
the mean density inside radius $r$ will therefore change as 
\begin{equation}\label{dp}
\hspace{-0.2truecm}\Delta \overline{\rho}_r =\ \Delta \left({{3 M(<r)}\over{4 \pi r^3}}\right)\ \propto - {\Delta r\over r^4} \propto 
- {\Delta E(r) \over r^2}\ .
\end{equation}

As equation (\ref{dp}) suggests, heating will cause the mass distribution to 
expand. Some of the material near the tidal radius will therefore cross this 
boundary and may be stripped away. 
Consequently, heating will accelerate mass loss.
Since $\Delta W_{\rm tid}(r)/r^2$ is independent of $r$, if we keep a running total of 
the eighteen terms in equation (\ref{dW2}), we can calculate the approximate
density change produced by tidal heating 
at some arbitrary radius $r$ as a function of time,
and then apply tidal stripping (eq.\ [\ref{overd}]) to the new, heated density
profile to determine how much mass is lost.  

This method for describing heating suffers from two limitations. First, 
we have used an average adiabatic correction for the system in equation
(\ref{adc}). The actual correction for an orbit of radius $r$ will depend 
on the orbital period at that radius, so the density change produced 
by heating will also depend weakly on radius. If we use a single scalar 
quantity to track $\Delta E /r^2$ for a given satellite, we will overestimate 
the heating experienced in its inner regions as a result. Second, we have 
assumed that the internal structure of the satellite doesn't change in the 
derivation of equation (\ref{dp}). On slow orbits, satellite structure may 
be partially re-virialized as the system relaxes between shocks, producing a 
tightly bound core which is resistant to subsequent tidal effects. In practice,
we expect these effects to be secondary, and to be partly masked by 
uncertainties in our choice of values for $\ln \Lambda$ and $\epsilon_{\rm h}$.

\section{Comparison with Numerical Results}

\subsection{Simulation Parameters}

To set the values of the three free parameters ($\ln \Lambda_{\rm h} $,
$\ln \Lambda_{\rm d} $ and $\epsilon_{\rm h} $) in our semi-analytic 
scheme and to test how well this prescription does in predicting the 
evolution of a subhalo moving inside a larger halo, we compare our
model results to 15 recent high-resolution simulations by VW
of the evolution of a single satellite within a larger halo containing 
a disk galaxy. Reproducing the results of these simulations 
offers a good test of our simplified description of merging, since each 
one follows the orbital evolution and mass loss history of the satellite 
in detail, and together they cover a range of different orbits and 
satellite densities. The satellites also have large masses 
and small orbital pericenters; thus if we can match these simulations 
reasonably well, we expect the agreement to be even better for the more
common case of small satellites orbiting at large distances in the halo. 

For the purpose of comparison, we evolved orbits in a static potential 
identical to the one adopted by VW, which consists 
of three components, a truncated isothermal halo with a core, a stellar 
bulge, and an exponential disk. The density profiles of the three components are:
\begin{eqnarray}
\rho_{\rm H}(r) & = & {{M_{\rm H}\alpha}\over{2\pi^{3/2}r_{\rm cut}}}{{\exp(-r^2/r_{\rm cut}^2)}\over{r^2 + \gamma^2}}\nonumber\\ 
\rho_{\rm B}(r) & = & {{M_{\rm B}}\over{2\pi}}{{a}\over{r(a + r)^3}}\nonumber\\ 
\rho_{\rm D}(r) & = & {{M_{\rm D}}\over{4\pi R_{\rm D}^2z_0}}\exp(-R/R_{\rm D})\,{\rm sech}^2(z/z_0)\nonumber
\end{eqnarray}
where the masses and scale lengths of the components are: 
\begin{eqnarray}
M_{\rm H}\hspace{-0.2cm}&=&\hspace{-0.2cm} 7.84 \times 10^{11}\msun, \gamma = 3.5\kpc, r_{\rm cut} = 84\kpc,\hfill\nonumber\\
M_{\rm B}\hspace{-0.2cm}&=&\hspace{-0.2cm} 1.87 \times 10^{10}\msun,\ \ a = 525 \pc,\hfill\nonumber\\
M_{\rm D}\hspace{-0.2cm}&=&\hspace{-0.2cm} 5.6 \times 10^{10}\msun, R_{\rm D} = 3.5\kpc,\ z_0 = 700\pc\ .\hfill\nonumber
\end{eqnarray}

The disk density used in the calculation of dynamical friction was smoothed 
in the vertical direction by two disk scale heights, as explained in 
section (2.2.1), to reflect
the finite size of the satellite, and we similarly smoothed the vertical 
component of the tidal field used in calculating heating by the disk.
The sum of the halo and bulge densities was used to calculate the other
friction term, since these components are kinematically similar.
For the velocity dispersions, we used: 
\begin{eqnarray}
\sigma_{\rm h} &=& (V^2_{\rm c,h} + V^2_{\rm c,b})^{1/2}/\sqrt{2} ,\hspace{0.7cm}\nonumber\\
{\rm and}\ \ \ \ \ \ \ \ {\phantom{\sigma_{\rm d}}}\nonumber\\
\sigma_{\rm d} &=& V_{\rm c,d}/\sqrt 2 = \sigma_{\rm o}\exp(-R/R_{\rm o})\hspace{0.7cm}\nonumber
\end{eqnarray}
where $V_{\rm c,h}$ , $V_{\rm c,b}$ and $V_{\rm c,d}$ 
are the circular velocities of the halo, bulge and disk respectively,
and we set $\sigma_{\rm o}$ to $143\kms$ and $R_{\rm o}$ to $7\kpc$ (or $2\,R_{\rm d}$),
based on the velocity dispersion of the disk measured by VW. 

Fifteen different orbits were simulated, with initial conditions 
corresponding to those of VW (see table 1). 
Our satellite models S1, S2, and S3 were also identical to those used by 
VW. These are King models, with core radii and initial concentrations 
appropriate to dwarf spheroidals (see table 2). 

\subsection{Results} 

\begin{figure}[ht!]
  \centerline{\psfig{figure=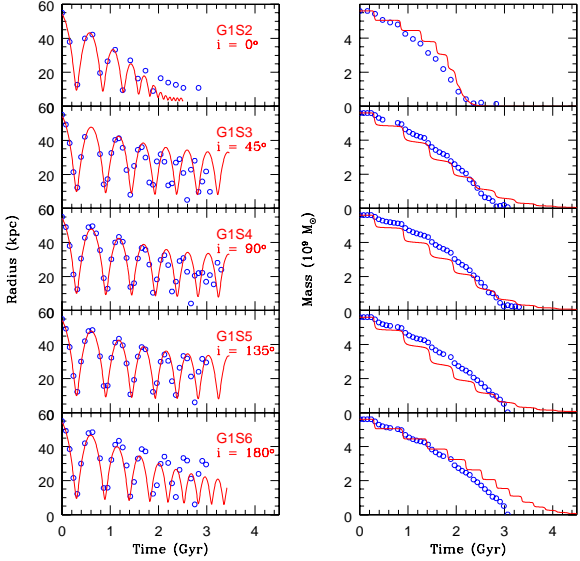,width=1.0\linewidth,clip=,angle=0}}
 \caption[]{ Orbits and mass loss histories for satellite
S1 (lines), compared with numerical results from Vel{\'a}zquez \& White (1999) 
(points). The parameter values used were ($\ln \Lambda_{\rm h}$, 
$\ln \Lambda_{\rm d}$, $\epsilon_{\rm h}$) = ( 2.4, 0.5, 3.0). \label{fig1}}
\end{figure}

\begin{figure}[ht!]
  \centerline{\psfig{figure=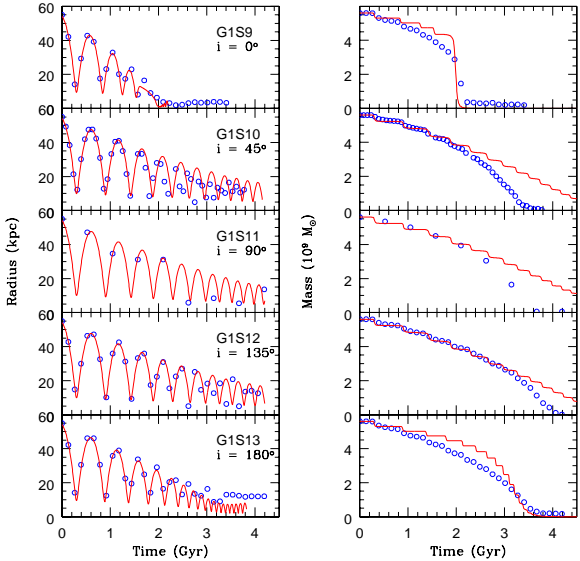,width=1.0\linewidth,clip=,angle=0}}
 \caption[]{As figure \ref{fig1}, 
but for a more concentrated satellite 
(Vel{\'a}zquez \& White model S2).\label{fig2} }
\end{figure}

Figure \ref{fig1} shows the evolution of satellite S1 on 
five orbits of different inclination with respect to the disk. 
The points are the results of the 
simulation, and the solid curves are the results from our semi-analytic model.
Figure \ref{fig2} shows similar results for the more concentrated satellite S2.
In each figure, the left-hand plots show the position of the
satellite versus time, while the right-hand plots show the mass.  
The angle $i$ indicated on the plots is the angle between the 
initial angular momentum vectors of the satellite and of the disk, 
so that orbits G1S2 and G1S9 are coplanar with the disk, and prograde
with respect to disk rotation,
while orbits G1S6 and G1S13 are coplanar and retrograde.  

The semi-analytic orbits were calculated using Coulomb logarithms 
of $\ln \Lambda_{\rm h} = 2.4 $ for the halo and $\ln \Lambda_{\rm d} = 0.5$ 
for the disk, which are in the range
predicted by the theoretical estimates mentioned in section 2
($\ln \Lambda_{\rm h}$ = 1.9--2.6 and $\ln \Lambda_{\rm d}$ = 0.6--1.3 or 
less, depending on the orbit and satellite properties).
The heating coefficient used, $\epsilon_{\rm h} = 3.0$, is also 
in the expected range. We see that for this choice
of parameter values, we obtain a very good match overall to the orbital decay
and mass loss in all ten cases.

Examining the orbital evolution in detail, we note that the semi-analytic
model matches the numerical results remarkably well, especially
given that we are using a single set of parameter values to fit results 
for three
different satellites and eight different sets of initial conditions. 
Our prescription
for dynamical friction reproduces the decay in the amplitude and 
period of the orbit, and the semi-analytic orbit remains in phase with the 
numerical results for as long as the mass loss is well matched, typically 
five or six orbital periods. Varying the Coulomb logarithm by 10--20\%
would produce a better match to some orbits, such as the retrograde,
coplanar orbit G1S6, but as mentioned previously, we prefer to find
a single set of values which fit all fifteen orbits reasonably well.
Varying the parameters by less than 10\%\ does not affect the results 
substantially.

Comparing the mass loss rates, we see that the semi-analytic model
gives an excellent estimate of the timescale for mass loss, and predicts 
the bound mass in the simulations to within 20\%, up to the point
where the satellite has lost most of its mass.  
Our model also reproduces the dependence of mass loss on the orientation
of the orbit for prograde and retrograde orbits in the disk, predicting
faster mass loss on prograde orbits. 
This appears to be mainly the result of the stronger dynamical friction
experienced by satellites in this case.
Orbits out of the plane of the disk show a weak dependence on inclination 
in the simulations. We reproduce this marginally, though the amplitude
of the effect is much smaller in our model than in the simulations.
This may indicate that dynamical friction from the disk is more
important than we predict for these orbits.

\begin{figure}[ht!]
  \centerline{\psfig{figure=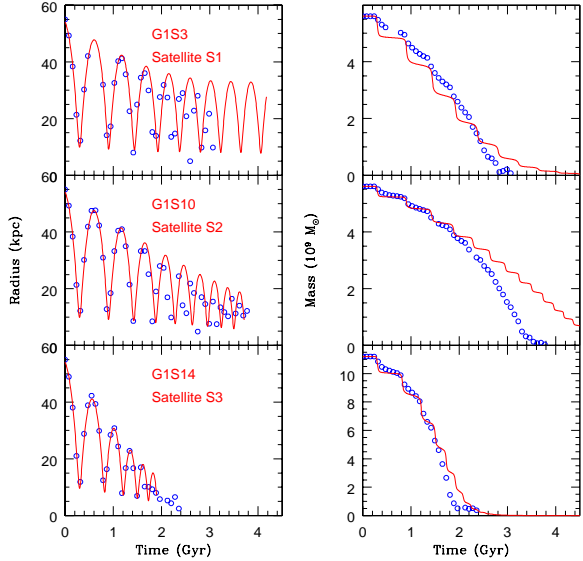,width=1.0\linewidth,clip=,angle=0}}
 \caption[]{Orbits and mass loss histories for three different
satellite models with the same initial orbital parameters.\label{fig3} }
\end{figure}

\begin{figure}[ht!]
  \centerline{\psfig{figure=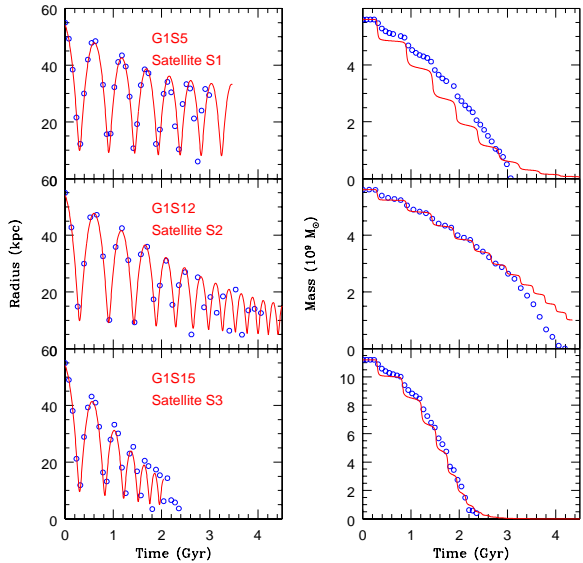,width=1.0\linewidth,clip=,angle=0}}
 \caption[]{As figure \ref{fig3}, for a retrograde 
orbit.\label{fig4} }
\end{figure}

Apart from considering a single satellite on a set of similar orbits at 
different inclinations, we can also consider different satellites on similar 
orbits or the same satellite on different orbits.  
Figures \ref{fig3} and \ref{fig4} show the results for three different 
satellites: 
the fiducial satellite (S1), a satellite which is more concentrated (S2),
and one which is more massive and more concentrated (S3), on prograde 
(Fig \ref{fig3}) 
and retrograde orbits (Fig \ref{fig4}), inclined by $45^\circ$.
For both prograde and retrograde orbits, the more massive satellite 
experiences more dynamical friction, falls in faster, and is disrupted. 
The more concentrated satellite retains its mass longer than S1,
despite having fallen 
further into the potential. The semi-analytic model accurately reproduces 
these trends, although for the more concentrated satellite the mass loss 
rates are a bit slow.   

\begin{figure}[ht!]
  \centerline{\psfig{figure=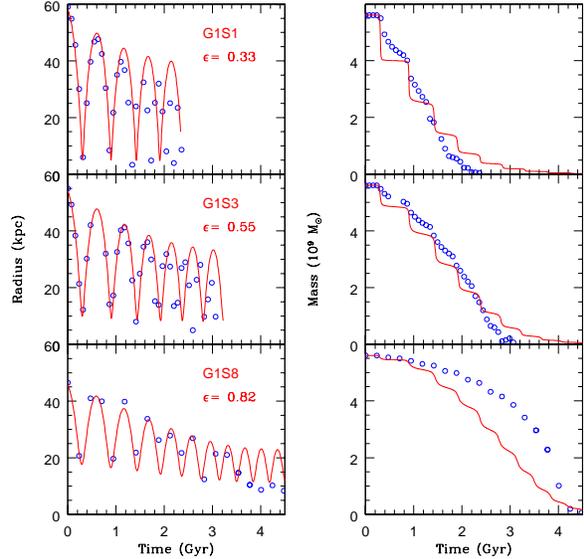,width=1.0\linewidth,clip=,angle=0}}
 \caption[]{Orbits and mass loss histories for satellite S1,
on orbits of different circularity. \label{fig5} }
\end{figure}

Dynamical friction, tidal limits, heating, and mass loss timescales
will also depend on the circularity of a satellite's orbit. Figure \ref{fig5}
shows results for the same satellite model (S1), on inclined orbits of 
three different eccentricities. Here again, we achieve an excellent match 
to the simulation results, reproducing the trend of faster mass loss for 
more eccentric orbits, and getting a good estimate of the disruption times 
for the three different orbits. 

Some systematic differences are apparent in the comparison between the 
numerical and semi-analytic results. The mass loss in the 
simulations is a bit smoother than in the semi-analytic model, 
showing less variation in rate at pericentric passage. We also
underestimate the mass loss rates for the more concentrated satellite, when
it is on orbits inclined with respect to the disk (G1S10--G1S12). Varying
the parameter values suggests that this may be due to a slight underestimate
of the dynamical friction in these cases, although our model for mass loss
is no doubt partly responsible for the mismatch. Some of the theoretical 
estimates for dynamical friction mentioned in section 2 do predict a
larger Coulomb logarithm for more concentrated satellites, so we may
be seeing evidence of this in our results. In the absence of more numerical 
results to confirm this dependence on concentration, however, we
will limit ourselves to a single set of values for the Coulomb logarithms.
Finally, our most circular orbit (G1S8) experiences slightly less mass loss 
than predicted. In this case, the characteristic timescales for the shocks
are very close to the internal orbital period of the satellite.  
Using a more restrictive definition of rapid shocks in this case produces 
results which match the numerical behaviour exactly. Here again though, 
we have insufficient numerical data to justify a general modification to 
our scheme.

We also note that the details of the numerical mass loss histories may depend 
on the precise definition of the bound mass used in interpreting the 
simulations. 
Investigating mass loss in detail would require a careful re-analysis 
of the simulations, given the importance of this and other purely numerical 
effects on the mass loss histories. VW show, for instance, that 
changing the resolution of a simulation by a factor of four can have an 
effect comparable to the discrepancy we see between the semi-analytic and 
numerical results (VW, figure 10). The softening lengths and time-stepping 
algorithms used could affect the numerical results to a similar degree. 
It is also intrinsically hard to separate out the 
effects of tidal heating and tidal stripping in the results of VW, since the 
mass loss timescales produced by the two are similar in their simulations. 
The fact that we find good agreement with their results over a range of orbits 
and for several satellite models, however, gives us some confidence in our 
description of these phenomena.

In summary, using a simple model of dynamical friction, tidal heating
and tidal mass loss, we can reproduce analytically the results of 
high-resolution $N$-body simulations of mergers, including accurate
timescales for satellite infall and disruption, with the correct dependence 
on satellite mass and concentration, and on orbital parameters. 
Our model has three free parameters --- $\ln \Lambda_{\rm h}$, 
$\ln \Lambda_{\rm d}$, and $\epsilon_{\rm h}$. 
Of the three, the results depend most strongly on $\ln \Lambda_{\rm h}$ and 
$\epsilon_{\rm h}$, with $\ln \Lambda_{\rm d}$ making a secondary contribution.
Requiring that our results match those of VW fixes the values of these 
parameters to within roughly 
10\%. The values we obtain fall within the range predicted by first-order
estimates of friction and heating.

\section{Discussion}
 
 \subsection{The importance of tidal heating}

\begin{figure}[ht!]
  \centerline{\psfig{figure=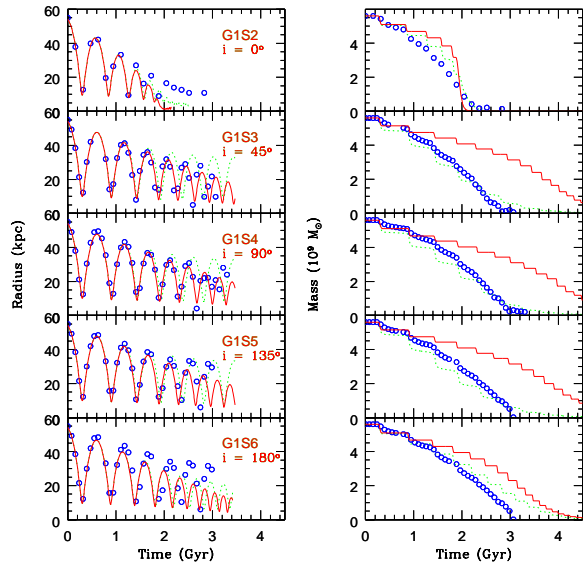,width=1.0\linewidth,clip=,angle=0}}
 \caption[]{A set of orbits calculated with and without heating
(dotted and solid lines respectively), compared with simulations (points).
The parameter values used were ($\ln \Lambda_{\rm h}$, 
$\ln \Lambda_{\rm d}$, $\epsilon_{\rm h}$)
= ( 2.4, 0.5, 3.0) and ( 2.4, 0.5, 0.0) for the heated and unheated cases,
 respectively.\label{fig6} }
\end{figure}

Since the processes underlying dynamical evolution are specified
explicitly in our model, it is possible to adjust them to test their relative
contribution to satellite evolution. In particular, we can test
the importance of tidal heating, often neglected in the study of
satellite dynamics, on our results. The solid curves in figure \ref{fig6} 
show several orbits calculated without tidal heating (solid curves), 
compared with the heated orbits (dotted curves), and the simulations 
(solid dots). We see that the overall effect of heating is to increase 
mass loss, which in turn reduces dynamical friction. In general, the 
simulation results are better matched by including heating, although 
the importance of heating varies with circularity and inclination.
On inclined orbits, satellites are strongly affected by heating, while its
effect on orbits in the plane of the disk is minor. This demonstrates 
that for the orbits we have considered, disk shocks dominate over bulge 
shocks as a source of heating.

If we consider a satellite to be disrupted when it has lost some large fraction
of its mass, the disruption times we measure for our satellites 
are up to 40\%\ shorter due to heating. One might expect an even stronger
effect, but we note, comparing the mass loss and orbital decay curves,
that dynamical friction conspires to reduce the difference between the 
mass loss times for the satellites and orbits considered here. 
In the no-heating runs, the satellites retain more of their mass initially
and therefore experience more dynamical friction. The satellite orbits 
decay faster as a result, and the satellites fall
into the center of the potential and are disrupted. This accounts
for the sharper cutoff to some of the mass loss curves. If the overall
timescales for disruption are similar with and without heating, 
this is partly an accident of the concentrations 
and masses of our satellites. A less massive satellite of similar
density, for instance, would have a much longer orbital decay time
but would lose mass through heating at about the same rate.
More concentrated satellites may resist heating almost completely, 
while less concentrated satellites may quickly be disrupted by it. 
Finally, heating will produce a quite different
distribution of stripped material from satellites, which is important
in considering halo substructure formed by satellite accretion and disruption.
Thus, heating cannot be neglected in studying minor mergers with semi-analytic
models.

 \subsection{The effect of the disk and the bulge}

The disk was shown above to have a strong effect on satellite evolution
by heating satellites on inclined orbits. Its presence should also
generate dynamical friction, particularly for prograde, coplanar orbits.
The effect of the bulge is less clear. Understanding the role of these
structures is important in relating small-scale simulations such as those
of VW to larger cosmological simulations, which do not yet include disks or bulges.
To test the effect of these components on satellite orbits, mass loss, and 
disruption times, and to determine any systematic trends affecting simulation 
results, we have run our model with one or both of these components removed.

Removing the bulge from the potential has little effect 
on the satellite orbits, acting only to decrease friction slightly.
We expect the disk to have a much greater effect, due to its greater mass, 
which produces more dynamical friction, and to its vertical density gradient, 
about ten times larger than that of the bulge or halo, which should 
contribute roughly 100 times more heating than the other components, for 
satellites crossing the disk plane. Running the model without a disk confirms 
that this is indeed the case.

\begin{figure}[ht!]
  \centerline{\psfig{figure=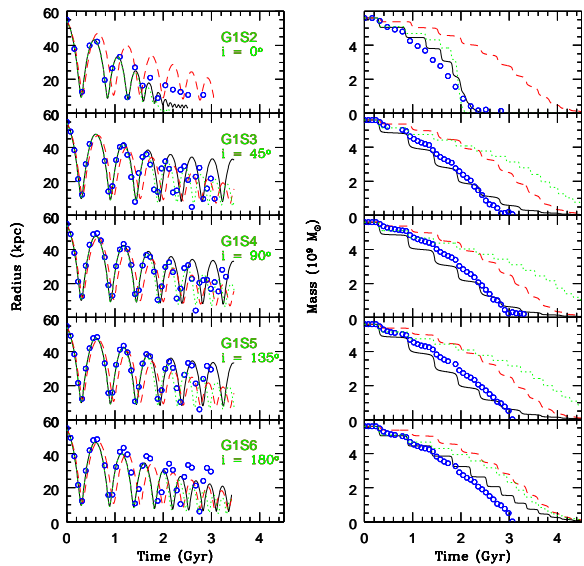,width=1.0\linewidth,clip=,angle=0}}
 \caption[]{A set of orbits calculated without a disk
component in the potential (dashed lines), compared with the no-heating case
(dotted lines), with the full model (solid lines) and with simulations 
(points). The parameter values
used were ($\ln \Lambda_{\rm h}$, $\ln \Lambda_{\rm d}$, 
$\epsilon_{\rm h}$) = ( 2.4, 0.5, 3.0 or 0.0).\label{fig7} }
\end{figure}

In figure \ref{fig7}, we show
satellite orbits G1S2--G1S9, recalculated in the same potential without
a disk (dashed curves), as well as the previous results for orbits in 
the presence of a disk, but with heating turned off (dotted curves), 
and with both a disk and heating (solid curves). When the disk is absent, 
we see that the dependence of orbital evolution on inclination vanishes, 
as expected. Furthermore, the initial mass loss rate is reduced, and satellites
fall further into the potential without losing as much mass. 
Turning off the disk or turning off heating produces similar results
for orbits G1S3--G1S5, indicating that the effect of the disk 
is mainly to heat satellites on inclined orbits.
In the prograde, coplanar orbit G1S2, on the other hand, the disk contributes 
mainly to dynamical friction. For this orbit, turning off heating does not 
change the results substantially, whereas turning off the disk does. 

Overall, we conclude that the presence of disk has an important effect 
on the evolution of satellites on orbits with pericenters of $20\kpc$ 
or less (about 6 scale radii). 
For the typical eccentricities seen in cosmological simulations 
(Ghigna et al.\ 1998), this suggests that orbits with apocenters of 
$120\kpc$ or more will be affected by the disk.
The effect of the disk on the disruption times we measure 
is to reduce them by 20--30\%, but as with heating, 
this difference could be much larger 
for satellites of different masses or concentrations. To produce realistic
distributions of galactic satellites by semi-analytic means,
(e.g.\ Bullock et al. 2000), one should therefore account for 
the effects of a disk in the model.

\subsection{Results for different satellite profiles}

Finally, the mass and concentration of a satellite can substantially
affect its dynamical evolution. One difficulty in comparing numerical studies 
of
disk heating through minor mergers is the fact that different authors have 
considered different satellite models in their simulations. In this section, 
we recalculate a few orbits
using some of the models that appear in the literature, to test the effect
of a satellite's internal structure on its orbital evolution.

\begin{figure}[ht!]
  \centerline{\psfig{figure=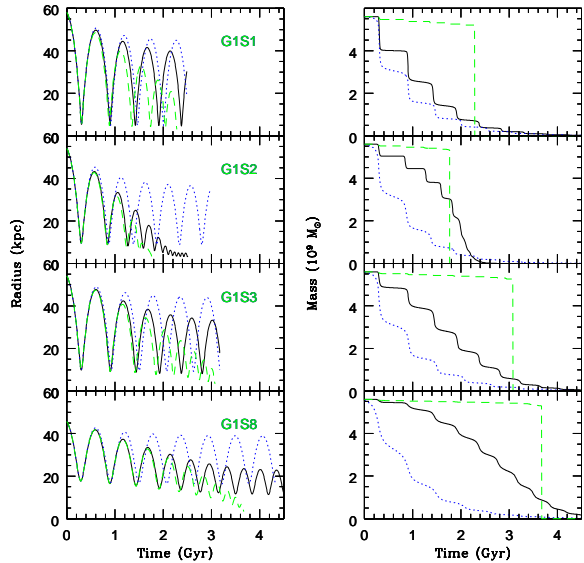,width=1.0\linewidth,clip=,angle=0}}
 \caption[]{A set of orbits calculated for various satellite
models. The solid lines are for Vel{\'a}zquez \& White's satellite S1,
the dotted line is the satellite of Huang and Carlberg,
and the short-dashed line is the more concentrated of 
T\'{o}th and Ostriker's two satellites. See text and table 2 for the details
of the models. The parameter values used were 
($\ln \Lambda_{\rm h}$, $\ln \Lambda_{\rm d}$, 
$\epsilon_{\rm h}$) = ( 2.4, 0.5, 3.0), as above.\label{fig8} }
\end{figure}

\begin{figure}[ht!]
  \centerline{\psfig{figure=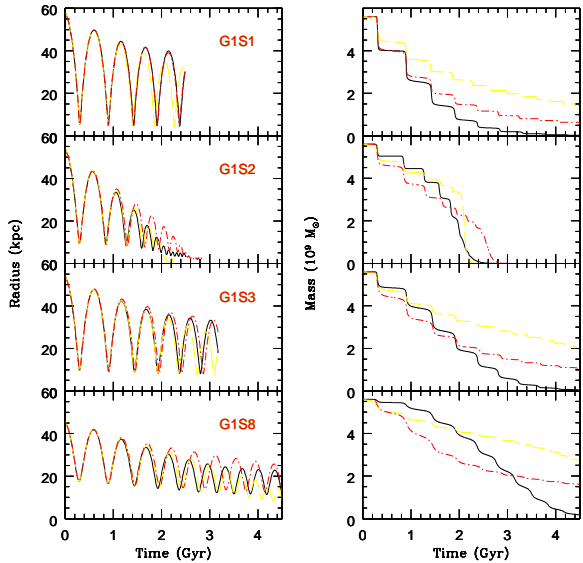,width=1.0\linewidth,clip=,angle=0}}
 \caption[]{As figure \ref{fig8}, for Vel{\'a}zquez \& White's satellite S1
(solid line),  the satellite of Walker, Mihos \& Hernquist (long-dashed line), 
and a satellite with an NFW profile (dot-dashed line). See table 2 
for the details of the models. The parameter values 
used were ($\ln \Lambda_{\rm h}$, $\ln \Lambda_{\rm d}$, 
$\epsilon_{\rm h}$) = ( 2.4, 0.5, 3.0), as above.\label{fig9} }
\end{figure}

Figures \ref{fig8} and \ref{fig9} show several orbits from the set used 
previously, 
recalculated for five different satellite models similar to those used in 
recent merger simulations. In figure \ref{fig8}, the solid curves are for 
the VW satellite S1, the dotted curves are for a more concentrated king model, 
similar to the one used by HC, and the short-dashed curves are for the 
highly concentrated satellite considered by TO.
In figure \ref{fig9}, the solid curves are for VW S1 as before, 
the long-dashed curves are for the satellite model used by WMH, and 
the dot-dashed curves are for a 
satellite with an NFW profile (Navarro et al.\ 1996, 1997) 
of comparable 
concentration. All of the satellite models have been given the same mass for 
purposes of comparison. Their density profiles and structural parameters are 
listed in table 2. 

We see from the mass loss curves the strong effect that the satellite model
has on dynamical evolution. The HC satellite is much more concentrated than 
S1, but has a similar core radius, and by implication an extended diffuse 
envelope, containing most of the satellite's mass. This diffuse material is 
stripped off early on in its orbit, leading to much slower orbital decay. The 
TO satellite behaves in the opposite way --- its mass is so tightly bound that 
it experiences almost no tidal stripping, falling directly into the center of 
the potential with little mass loss. Finally the WMH satellite looses a 
comparable amount of mass to S1 initially, but its overall mass loss history 
produces much slower orbital decay. The same is true for the satellite with 
an NFW profile. 

These results are consistent with those of the original studies. HC found 
that their satellites were stripped of most of their mass before they hit the 
disk, while TO's more concentrated satellite retained 90\%\ of its mass as it 
fell in on a circular orbit to a final radius of $4 \kpc$. WMH found that a 
fair amount of mass was stripped off in the outer regions of the halo-disk 
system, but that the satellite managed to carry as much as half of its mass 
into the central few kpc. Here we have compared satellites which differ only 
in density profile, on the same orbit in the same potential. Since the authors 
mentioned above consider different satellite masses, different orbits, and 
different forms of the galactic potential in their simulations, the 
semi-analytic mass loss rates shown in figures \ref{fig8} and \ref{fig9} 
will differ in detail from their numerical results, but we certainly reproduce 
all of the trends mentioned.

\begin{figure}[ht!]
  \centerline{\psfig{figure=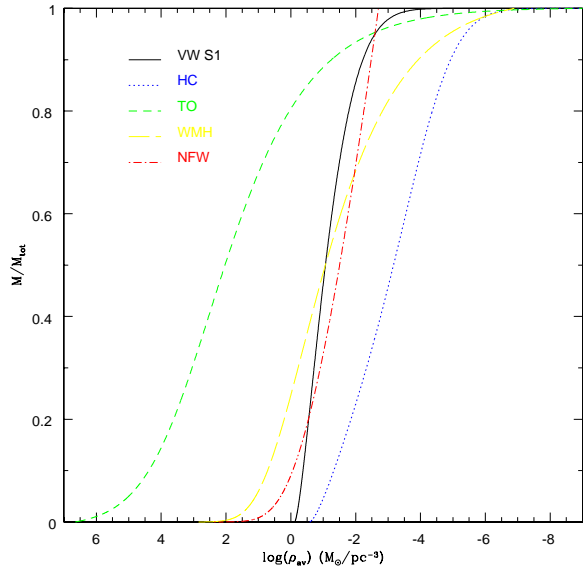,width=1.0\linewidth,clip=,angle=0}}
 \caption[]{A plot of the fraction of mass within a given 
radius, as a function of mean density 
within this radius, for the various satellite models.
Line styles are as in figures \ref{fig8} \& \ref{fig9}.\label{fig10} }
\end{figure}

One way of understanding these different mass loss rates is in terms of 
the fraction of a satellite's mass that lies within a given mean density 
contour. This structural property is related to the concentration of a 
satellite, and to its density profile. Figure \ref{fig10} shows the mass 
fraction as a function of density, plotted for the different profiles 
considered here. We see that all the mass in the HC model is at lower 
densities than the VW model S1. Almost all the mass in the TO profile, on 
the other hand, is at densities much higher than S1, and higher than the 
central density of the main galaxy (which is roughly $1\msun\pc^{-3}$). 
For the WMH profile, half the mass is at densities lower than VW model S1, 
but the core of the satellite is at higher densities. These different profiles
lead to different mass loss rates throughout the orbit of the satellite, and 
as a result, very different dynamical histories. 

\begin{figure}[ht!]
  \centerline{\psfig{figure=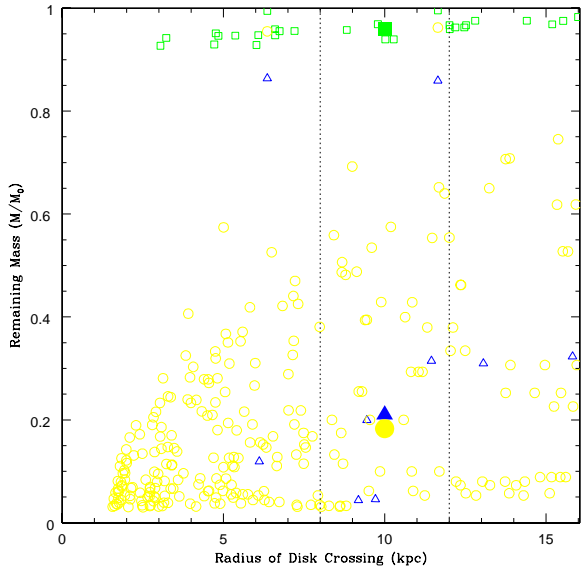,width=1.0\linewidth,clip=,angle=0}}
 \caption[]{Disk crossings in orbits G1S1, G1S3 and G1S8. The fraction 
of mass remaining bound to the satellite is plotted versus 
the radius at which it crosses the disk. The symbols indicate the satellite 
models of TO (squares), HC (triangles), and WMH (circles).
The bold symbols indicate the average mass remaining for all disk
crossings between 8 and 12 \kpc.\label{fig11} }
\end{figure}

In particular, the amount of mass a satellite looses in the outer part of 
its orbit, before it hits the disk, can vary tremendously from one model to 
another. Figure \ref{fig11} shows all the disk crossings recorded in three 
orbits (G1S1, G1S3 and G1S8), plotted in terms of the fraction of the 
satellite's original mass that is still bound to it at that point, versus 
the radius at which it crosses the disk. The different symbols indicate the 
satellite models of TO (squares), HC (triangles), and WMH (circles). We see 
that while the TO satellite crosses the disk many times with almost all of 
its mass intact, less dense satellites such as that of HC have been stripped 
of most of their mass after a few orbits. The filled symbols in figure 
\ref{fig11} indicate the average mass fraction for all disk crossings between
8 and 12\kpc. While TO's satellite encounters the disk at this radius
with 96\%\ of its mass intact on average, the satellites of HC and WMH
have only 20\%\ of their mass intact at this point. Given that TO saw more 
disk heating in their study than HC or WMH, these results suggest that 
heating may be simply related to the mass of material accreted by the disk, 
once tidal stripping has been taken into account. We shall investigate this 
possibility in detail in a subsequent paper. In any case, when studying minor 
mergers, accretion and disk heating, it is clearly important to use 
cosmologically motivated density profiles, and to consider how different 
satellite models may affect the final results.

\section{Conclusion}

While there has been much progress recently in the understanding of how
structure forms in dark matter on cluster, group and galaxy scales, 
the dynamical evolution of halos and subhalos is still not very well 
understood. Numerical simulations typically lack the resolution and 
statistics to follow the formation and evolution of structure across the
range of scales involved, and much of the underlying physics remains
uncertain. Several observed features of galaxies, such as thin disks, 
seem difficult to explain in current 
hierarchical models. This may partly be the result of the computational
limitations restricting current numerical studies, or it may imply a 
genuine problem with the underlying cosmological models. To explore 
the parameter space relevant to these issues requires a method which is
faster and less computationally intensive than numerical simulation. To 
this end, we have developed a simple model of the dynamical evolution of 
substructure on galactic scales.

Our model follows the dynamics of individual subhalos numerically,
but accounts for dynamical friction, tidal mass loss and tidal heating
using analytic expressions from Chandrasekhar (1943), King (1962),
Gnedin \& Ostriker (1999) and Gnedin et al.\ (1999). 
We calibrate the model by comparison with 
fully numerical simulations. In particular, we find that we can reproduce
the results of the most recent set of high-resolution simulations of 
satellite infall by Vel{\'a}zquez \& White (1999), and that matching
these simulations sets our three free parameters to within roughly 
10\%. The values we obtain are all in the range predicted by first-order
estimates of friction and heating.

Varying the shape of background potential, the amount of tidal heating
and the density profile of the satellite in our model,
we can start to extract from these simulations the factors
contributing to mass loss and orbital decay. In general, tidal heating 
increases the mass loss and orbital decay times of our satellites 
substantially, although these effects are partly
masked by dynamical friction for the satellites and orbits we consider. 
We find in particular that the presence of a thin disk strongly affects 
evolution of objects in the inner regions of the galaxy,
while the presence of a central bulge has little effect. 
For the orbital eccentricities typically seen in cosmological 
simulations, satellites on orbits with apocenters of $120\kpc$ or more 
will pass through the disk repeatedly within a Hubble time, so it is important 
to consider its effects when studying galactic satellites. 
The overall evolution of a satellite is sensitive to its density profile. 
In the tidal truncation approximation, for instance, the satellite's mass 
loss history is determined by its mass fraction as a function of mean density. 
This dependence may explain some of the discrepancies found between 
different simulations of disk heating through satellite infall.

Given the excellent overall match to simulations of minor mergers that we 
achieve using a simple, physically motivated model of satellite dynamics, 
we can go on to consider the evolution of the large numbers of subhalos that 
a galactic halo will accrete over its lifetime. In the next paper in this 
series, we describe how to construct the mass accretion history of a halo 
from a merger tree, and use it as the input to our model of dynamical 
evolution. We shall then apply this method to several outstanding problems 
in galaxy formation, notably the question of disk survival in hierarchical 
models.

\acknowledgements

The authors thank H.\ Vel{\'a}zquez and S.\ White for providing 
the data from their simulations. We also wish to thank J.\ Barnes, 
J.\ Navarro, S.\ White and R.\ Wyse for helpful discussions. 
JET gratefully acknowledges the support of a postgraduate scholarship from the 
Natural Sciences \& Engineering Research Council of Canada (NSERC) during the 
course of this work.
AB gratefully acknowledges the kind hospitality shown to him by the Institute
for Theoretical Physics (ITP) during the course of the Galaxy Formation 
workshop (January-April 2000) where some of this work described in this paper 
was carried out. This research has been partly supported by the 
National Science Foundation Grant No. PHY94-07194 to ITP as well as by 
operating grants to AB from
the Natural Science and Engineering Research Council of Canada (NSERC).

\clearpage




\begin{center}
\begin{deluxetable}{lcrccc}
\label {table1}
\tablewidth{20pc}
\tablecaption{Summary of Simulations from Vel{\'a}zquez \& White (1999)}
\tablehead{
\colhead{Name}           & \colhead{Satellite}      &
\colhead{$\theta_i$}           & \colhead{$\epsilon_J$}      &
\colhead{$r_{\rm p}$}           & \colhead{$r_{\rm a}$}      \\
\colhead{}           & \colhead{model}      &
\colhead{}           & \colhead{}      &
\colhead{(kpc)}           & \colhead{(kpc)}}
\startdata
G1S1  & S1  &  $ 45^\circ$ & $0.33$ &  $5.25$ & $59.0$ \\
G1S2  & S1  &  $  0^\circ$ & $0.55$ &  $10.5$ & $55.0$ \\
G1S3  & S1  &  $ 45^\circ$ & $0.55$ &  $10.5$ & $55.0$ \\
G1S4  & S1  &  $ 90^\circ$ & $0.55$ &  $10.5$ & $55.0$ \\
G1S5  & S1  &  $135^\circ$ & $0.55$ &  $10.5$ & $55.0$ \\ 
G1S6  & S1  &  $180^\circ$ & $0.55$ &  $10.5$ & $55.0$ \\
G1S7  & S1  &  $  0^\circ$ & $0.82$ &  $21.0$ & $46.5$ \\
G1S8  & S1  &  $ 45^\circ$ & $0.82$ &  $21.0$ & $46.5$ \\
G1S9  & S2  &  $  0^\circ$ & $0.55$ &  $10.5$ & $55.0$ \\
G1S10 & S2  &  $ 45^\circ$ & $0.55$ &  $10.5$ & $55.0$ \\
G1S11 & S2  &  $ 90^\circ$ & $0.55$ &  $10.5$ & $55.0$ \\
G1S12 & S2  &  $135^\circ$ & $0.55$ &  $10.5$ & $55.0$ \\
G1S13 & S2  &  $180^\circ$ & $0.55$ &  $10.5$ & $55.0$ \\
G1S14 & S3  &  $ 45^\circ$ & $0.55$ &  $10.5$ & $55.0$ \\ 
G1S15 & S3  &  $135^\circ$ & $0.55$ &  $10.5$ & $55.0$ \\
\enddata
\tablecomments{
$\theta_i$ refers to the angle between the initial angular momentum vector 
of
the satellite and the angular momentum vector of the disc. The circularity 
$\epsilon_J = J/J_{\rm c}$, where $J$ is the initial angular momentum
of the orbit and $J_{\rm c}$ is the angular momentum of a cicular orbit
with the same energy. $r_{\rm p}$ and $r_{\rm a}$ are 
the initial pericentric and apocentric radii of the orbit, respectively. 
}
\end{deluxetable}
\end{center}

\begin{deluxetable}{lccccc}
\label {table2}
\tablewidth{30pc}
\tablecaption{Satellite Models}
\tablehead{
\colhead{Satellite}           & \colhead{Model}      &
\colhead{Density Profile}     & \colhead{Mass}&
\colhead{$r_c$}               & \colhead{$r_t$}      \\
\colhead{}           & \colhead{}      &
\colhead{}     & \colhead{($\msun$)}&
\colhead{(pc)}               & \colhead{(kpc)}      
}
\startdata
VW S1 & King &$\simeq A/(r^2 + r^2_c)$&$5.6\phd \times 10^{9\phd}$&$1000$&$\phd 6.3$\\
VW S2 &$\ ^\prime \ ^\prime$&out to $r_t$&$5.6\phd \times 10^{9\phd}$&$500$&$\phd 6.3$\\
VW S3 &$\ ^\prime \ ^\prime$&$\ ^\prime \ ^\prime$&$1.12 \times 10^{10}$&$850$&$\phd 8.5$\\
HC          &$\ ^\prime \ ^\prime$&$\ ^\prime \ ^\prime$&$5.6\phd \times 10^{9\phd}$&$812$&$55.3$\\
TO          & Jaffee    &$A/r^2(r + r_c)^2$&$5.6\phd \times 10^{9\phd}$&$100$&$47.2$\\
WMH&Hernquist&$A/r(r + r_c)^{3}$&$5.6\phd \times 10^{9\phd}$&$525$&$45.5$\\
NFW   & NFW       &$A/r(r + r_c)^2$&$5.6\phd \times 10^{9\phd}$&$500$&$\phd 5.0$\\
\enddata
\end{deluxetable}


\clearpage


\begin{thebibliography}{}

\bibitem[Barnes (1998)]{b98} Barnes, J. E. 1998, 
Galaxies: Interactions and Induced Star Formation, Saas-Fee Advanced Course 
26, 275 

\bibitem[Benson, et al.\ (1999)]{b99} Benson, A. J., Pearce, F. R., Frenk, 
C. S., Baugh, C. M. \& Jenkins, A. 2000, \mnras, submitted (astro-ph/9912220)

\bibitem[Binney \& Tremaine (1987)]{bt87} Binney, J. \& Tremaine, S. 1987, 
Galactic Dynamics (Princeton: Princeton University Press)

\bibitem[Bontekoe \& van Albada (1987)]{ba87} Bontekoe, T. 
R. \& van Albada, T. S. 1987, \mnras, 224, 349 

\bibitem[Bullock, Kravtsov \& Weinberg(2000)]{bk00} Bullock, 
J.\ S., Kravtsov, A.\ V.\ \& Weinberg, D.\ H.\ 2000, \apj, 539, 517 

\bibitem[Chandrasekhar (1943)]{c43} Chandrasekhar, S. 1943, 
\apj, 97, 255 

\bibitem[Cole, et al.\ (1994)]{ca94} Cole, S., Aragon-Salamanca, A., 
Frenk, C. S., Navarro, J. F. \& Zepf, S. E. 1994, \mnras, 271, 781 

\bibitem[Colpi, Mayer \& Governato (1999)]{cm99} Colpi, M., 
Mayer, L. \& Governato, F. 1999, \apj, 525, 720 

\bibitem[Cora, Muzzio \& Vergne (1997)]{cm97} Cora, S. A., 
Muzzio, J. C. \& Vergne, M. M. 1997, \mnras, 289, 253 

\bibitem[Dalcanton, Spergel \& Summers (1997)]{ds97} 
Dalcanton, J. J., Spergel, D. N. \& Summers, F. J. 1997, \apj, 482, 659 

\bibitem[Diaferio et al.\ (2000)]{di00} Diaferio, A., Kauffmann, G., 
Balogh, M. L., White, S. D. M., Schade, D. \& Ellingson, E. 2000, 
\mnras, submitted (astro-ph/0005485)


\bibitem[Dom\'{\i}nguez-Tenreiro \& G\'{o}mez-Flechoso 1998]{dg98}
Dom\'{\i}nguez-Tenreiro, R., \& G\'{o}mez-Flechoso, M. A. 1998, \mnras, 294, 465 

\bibitem[Fukushige \& Makino 2000]{fm00}Fukushige, T. \& Makino, J 2000, 
\apj, submitted (astro-ph/0008104)

\bibitem[Ghigna, et al.\ (1998)]{gm98} Ghigna, S., Moore, B. 
, Governato, F., Lake, G., Quinn, T. \& Stadel, J. 1998, \mnras, 300, 
146 

\bibitem[Ghigna, et al.\ (1999)]{gm99} Ghigna, S., Moore, B. 
, Governato, F., Lake, G., Quinn, T. \& Stadel, J. 1999, \apj, in press
(astro-ph/9910166)

\bibitem[Gnedin \& Ostriker (1997)]{go97} Gnedin, O. Y. \& 
Ostriker, J. P. 1997, \apj, 474, 223 

\bibitem[Gnedin \& Ostriker (1999)]{go99} Gnedin, O. Y. \& 
Ostriker, J. P. 1999, \apj, 513, 626 

\bibitem[Gnedin, Hernquist \& Ostriker (1999)]{gh99} Gnedin, 
O. Y., Hernquist, L. \& Ostriker, J. P. 1999, \apj, 514, 109 

\bibitem[Governato, et al.\ (1998)]{gb98} Governato, F., 
Baugh, C. M., Frenk, C. S., Cole, S., Lacey, C. G., Quinn, T. \& Stadel, J. 
1998, \nat, 392, 359 

\bibitem[Governato, et al.\ (1999)]{gb99} Governato, F., 
Babul, A., Quinn, T., Tozzi, P., Baugh, C. M., Katz, N. \& Lake, G. 1999, 
\mnras, 307, 949

\bibitem[Holmberg (1969)]{ho69} Holmberg, E. 1969, Arkiv. Astr., 5, 305

\bibitem[Huang \& Carlberg (1997)]{hc97} Huang, S. \& 
Carlberg, R. G. 1997, \apj, 480, 503 

\bibitem[Innanen, Harris \& Webbink (1983)]{ih83}
Innanen, K. A., Harris, W. E. \& Webbink, R. F. 1983, \aj, 88, 338

\bibitem[Jenkins et al.\ (1998)]{jf98} Jenkins, A., Frenk, C. S., 
Pearce, F. R., Thomas, P. A., Colberg, J. M., White, S. D. M., 
Couchman, H. M. P., Peacock, J. A., Efstathiou, G. 
\& Nelson, A. H. 1998, \apj, 499, 20 

\bibitem[Jing \& Suto (1998)]{js98} Jing, Y. P. \& Suto, 
Y. 1998, \apjl, 494, L5 

\bibitem[Jing \& Suto (2000)]{js00} Jing, Y. P. \& Suto, 
Y. 2000, \apjl, 529, L69 

\bibitem[Kauffmann \& White (1993)]{kw93} 
Kauffmann, G. \& White, S. D. M. 1993, \mnras, 261, 921 

\bibitem[Kauffmann, White \& Guiderdoni (1993)]{kwg3} 
Kauffmann, G., White, S. D. M. \& Guiderdoni, B. 1993, \mnras, 264, 201 

\bibitem[Kauffmann, Colberg, Diaferio \& White (1999)]{kc99} 
Kauffmann, G., Colberg, J. M., Diaferio, A. \& White, S. D. M. 1999, 
\mnras, 303, 188 

\bibitem[King (1962)]{ki62} King, I. 1962, \aj, 67, 471 

\bibitem[Klypin, Gottl\"ober, Kravtsov \& Khokhlov (1999a)]{kg99} 
Klypin, A., Gottl\"ober, S., Kravtsov, 
A. V. \& Khokhlov, A. M. 1999a, \apj, 516, 530 

\bibitem[Klypin, Kravtsov, Valenzuela \& Prada (1999b)]{kk99} 
Klypin, A., Kravtsov, A. V., Valenzuela, O. \& Prada, F. 1999b, \apj, 
522, 82 

\bibitem[Knebe, Kravtsov, Gottl{\"o}ber \& Klypin(2000)]{kk00} 
Knebe, A., Kravtsov, A.\ V., Gottl{\"o}ber, S.\ \& Klypin, A.\ A.\ 2000, \mnras, 317, 630 

\bibitem[Kundi\'{c} \& Ostriker (1995)]{ko95} Kundi\'{c}, T. \& 
Ostriker, J. P. 1995, \apj, 438, 702 

\bibitem[Lacey \& Cole (1993)]{lc93} Lacey, C. \& Cole, S. 1993, \mnras, 262, 627

\bibitem[Lewis et al.(2000)]{lb00} Lewis, G. F., Babul, A., 
Katz, N., Quinn, T., Hernquist, L. \& Weinberg, D. H. 2000, \apj, 536, 
623 

\bibitem[Maoz(1993)]{ma93} Maoz, E. 1993, \mnras, 263, 75 

\bibitem[Mo, Mao \& White (1998)]{mm98} Mo, H. J., Mao, S. 
\& White, S. D. M. 1998, \mnras, 295, 319 

\bibitem[Moore, et al.\ (1999)]{mg99} Moore, B., Ghigna, S., Governato, F., 
Lake, G., Quinn, T., Stadel, J. \& Tozzi, P. 1999, \apjl, 524, L19 

\bibitem[Naab, Burkert \& Hernquist (1999)]{nb99} Naab, T., 
Burkert, A. \& Hernquist, L. 1999, \apjl, 523, L133 

\bibitem[Navarro, Frenk \& White (1994)]{nf94} Navarro, J., Frenk, C. S., 
White, S. D. M. 1996, \mnras, 267, L1

\bibitem[Navarro, Frenk \& White (1996)]{nf96} Navarro, J., Frenk, C. S., 
White, S. D. M. 1996, \apj, 462, 563

\bibitem[Navarro, Frenk \& White (1997)]{nf97} Navarro, J., Frenk, C. S., 
White, S. D. M. 1996, \apj, 490, 493

\bibitem[Navarro \& Steinmetz(2000)]{ns00} Navarro, J.\ F.\ 
\& Steinmetz, M.\ 2000, \apj, 538, 477 

\bibitem[Okamoto \& Habe (1999)]{oh99} Okamoto, T. \& Habe, 
A. 1999, \apj, 516, 591 

\bibitem[Ostriker, Spitzer \& Chevalier(1972)]{1972ApJ...176L..51O} 
Ostriker, J. P., Spitzer, L. J. \& Chevalier, R. A. 1972, \apjl, 176, 
L51 

\bibitem[Pearce, et al.\ (1999)]{pj99} Pearce, F. R., Jenkins, A., 
Frenk, C. S., Colberg, J. M., White, S. D. M., Thomas, P. A., 
Couchman, H. M. P., Peacock, J. A. \& Efstathiou, G. 1999, \apjl, 521, L99 

\bibitem[Press \& Schechter (1974)]{ps74} Press, W. H. \& 
Schechter, P. 1974, \apj, 187, 425 

\bibitem[Quinn, Hernquist \& Fullagar (1993)]{qh93} Quinn, 
P. J., Hernquist, L. \& Fullagar, D. P. 1993, \apj, 403, 74 

\bibitem[Quinn \& Goodman (1986)]{qg86} Quinn, P. J. \& 
Goodman, J. 1986, \apj, 309, 472 

\bibitem[Sellwood, Nelson \& Tremaine (1998)]{sn98} 
Sellwood, J. A., Nelson, R. W. \& Tremaine, S. 1998, \apj, 506, 590 

\bibitem[Sigad et al.\ (2000)]{sk00} Sigad, Y., Kolatt, T. S., Bullock, 
J. S., Kravtsov, A. V., Klypin, A. A., Primack, J. R. \& Dekel, A.2000, 
\mnras, submitted (astro-ph/0005323)

\bibitem[Somerville \& Kolatt (1999)]{sk99} Somerville, R. 
S. \& Kolatt, T. S. 1999, \mnras, 305, 1 

\bibitem[Somerville \& Primack (1999)]{sp99} Somerville, R. 
S. \& Primack, J. R. 1999, \mnras, 310, 1087 

\bibitem[Spitzer (1987)]{s87} Spitzer, L. J. 1987,
Dynamical Evolution of Globular Clusters 
(Princeton: Princeton University Press)

\bibitem[T\'{o}th \& Ostriker (1992)]{to92} T\'{o}th, G. \& 
Ostriker, J. P. 1992, \apj, 389, 5 

\bibitem[van den Bosch, Lewis, Lake \& Stadel (1999)]{vl99} 
van den Bosch, F. C., Lewis, G. F., Lake, G. \& Stadel, J. 1999, 
\apj, 515, 50 

\bibitem[Vel{\'a}zquez \& White (1999)]{vw99} 
Vel{\'a}zquez, H. \& White, S. D. M. 1999, \mnras, 304, 254 

\bibitem[von Hoerner (1957)]{v57}von Hoerner, S. 1957, \apj, 125, 451

\bibitem[Walker, Mihos \& Hernquist (1996)]{wm96} Walker, I. 
R., Mihos, J. C. \& Hernquist, L. 1996, \apj, 460, 121 

\bibitem[Weinberg (1986)]{w86} Weinberg, M. D. 1986, \apj, 300, 93 

\bibitem[Weinberg (1995)]{w95} Weinberg, M. D. 1995, \apj, 455, L31 

\bibitem[Weinberg (1998)]{w98} Weinberg, M. D. 1998, \mnras, 299, 499 

\bibitem[White(1976)]{wh76} White, S. D. M. 1976, \mnras, 
174, 467 

\bibitem[Yoshikawa, Jing \& Suto(2000)]{2000ApJ...535..593Y} Yoshikawa, K., 
Jing, Y.\ P.\ \& Suto, Y.\ 2000, \apj, 535, 593 

\end{thebibliography}
\end{document}